\documentstyle[aps]{revtex}

\begin{document}
\title{ANKARA\ UNIVERSITY FACULTY\ OF\ SCIENCES\hspace{1.0in}\hspace{1.0in}
DEPARTMENTS OF PHYSICS AND ENGINEERING\ PHYSICS\\
\smallskip\ \\
\hspace{1.0in}\hspace{1.0in}\hspace{1.0in} \hspace{1.0in}\hspace{1.0in}%
AU-HEP-98/01\\
\hspace{1.0in}\hspace{1.0in}\hspace{1.0in}\hspace{1.0in}\hspace{1.0in}15
January 1998\\
\vspace{1.0in}New Tool for ''Old'' Nuclear Physics: FEL $\gamma $ - Nucleus
Colliders}
\author{H. Akta\c {s}$^a$, N. B\"{u}get$^b$, A. K. \c {C}ift\c {c}i$^a$, N. Meri\c {c%
}$^a$, S. Sultansoy$^{b,c}$, \"{O}. Yava\c {s}$^b$}
\address{$^a$Physics Department, Faculty of Sciences, Ankara University, 06100\\
Tando\u {g}an, Ankara, TURKEY\\
$^b$Department of Engineering Physics, Faculty of Sciences, Ankara\\
University, 06100 Tando\u {g}an, Ankara, TURKEY \\
$^c$Institute of Physics, Academy of Sciences, H.Cavid avenue 143, Baku,\\
AZERBAIJAN\\
\vspace{1.0in}}
\maketitle

\begin{abstract}
A new method has been considered to investigate the scattering reactions
with photons on fully ionized nuclei. To induce $\gamma $-nucleus collisions
a free electron laser and a heavy ion synchrotron have been considered
bringing them together. Main parameters of the collider (especially LHC and
HERA for the acceleration of nuclei) have been estimated. Rough calculations
have also been made for the cross sections of the excited nuclei production.
Finally, some design problems involving the collider have been considered.

\break 
\end{abstract}

\section{Introduction}

Linac-ring type $ep$ and $\gamma p$ colliders have been proposed recently as
the fourth way to reach energies in the TeV scale \cite{ankara}. If future
linear $e^{+}e^{-}$ colliders (or special $e$-linacs) are constructed near
the proton rings of HERA ($ep$ collider, Germany) and FNAL ($\overline{p}p$
collider, USA) or LHC ($pp$ collider, CERN) under construction, a number of
additional opportunities will arise. For example \cite{sultansoy1},

\smallskip\ 

LHC$\otimes $TESLA = LHC$\oplus $TESLA

\qquad \qquad \qquad \qquad $\oplus $ TeV scale $ep$ collider \cite{tww}\cite
{acs}\cite{bd}\cite{bc...}

\qquad \qquad \qquad \qquad $\oplus $ TeV scale $\gamma p$ collider \cite
{ictp}\cite{csty}\cite{bc...}

\qquad \qquad \qquad \qquad $\oplus $ Multi-TeV scale $e-nucleus$ and $%
\gamma -nucleus$ colliders \cite{ichep}\cite{bc...}

\qquad \qquad \qquad \qquad $\oplus $ FEL $\gamma -nucleus$ collider

\medskip\ \ $\qquad $

In this paper we deal with the last collider. As known, TESLA can operate as
Free Electron Laser (FEL) in the X-ray region \cite{FEL}. Collision of the
FEL beam with the bunch of nuclei may give a unique possibility to
investigate ''old'' nuclear phenomena in rather unusual conditions. In fact,
keV energy FEL photons will be seen as a ''laser'' beam in the MeV energy
range in the rest frame of the nucleus. Earlier investigations on this
subject involve bremsstrahlung photons having continuous energy spectrum.
Moreover, since the accelerated nuclei are fully ionized, the background
which may be induced by the low-shell electrons will be eliminated. This
option needs further investigations from both the accelerator and the
nuclear physics point of view.

In section II general considerations of the proposed collider have been
presented. Luminosities of $\gamma $-nucleus collisions have been estimated
in section III. Next, in order to illustrate the physics search capacity of
the proposed machines two proceses have been given. Finally, some concluding
remarks have been presented in section V.

\section{General Considerations}

The main elements of the proposed machine are illustrated in Fig.1. The $%
\gamma $-beam produced by the Free Electron Laser based on TESLA (or a
special $e$-linac) collides with the nucleus beam from HERA\ or LHC. In the
nucleus rest frame the energy of FEL photon is multiplied by the Lorentz
factor $\gamma _{N\text{ }}$ of the nucleus ($\gamma _{N\text{ }}\simeq 500$
for HERA and $\gamma _{N\text{ }}\simeq 3000$ for LHC). The FEL photon
wavelength is given by \cite{FEL} 
\begin{equation}
\lambda =\frac{l_u}{2\gamma _e^2}\left( 1+\frac{k^2}2\right) ,\text{ }
\end{equation}
where $l_u$ is the period length of an undulator, $\gamma _e=E_e/m_e$ is the
Lorentz factor of the electron ($E_e$ and $m_e$ being the energy and mass of
the electron), $k=eB_ul_u/2\pi m_e$ is the undulator parameter in which $e=%
\sqrt{4\pi \alpha }$ ($\alpha $ being the fine structure constant) and $B_u$
is the peak value of the magnetic field in the undulator. For illustration,
let us use the design parameters \cite{FEL} of the TESLA Test Facility (TTF)
FEL, which are presented in Table 1. The nucleus will ''see'' TTF FEL beam
as photons with an energy of 96.5 keV at HERA and 579 keV at LHC. Last
values may be adjusted by changing $E_e$, $B_u$, $l_u$ or $\gamma _N$. With
the nesessary adjustment the energy region can be made suitable for the
investigation of the electromagnetic excitations of the nucleus.

The excited nucleus will turn to the ground state at a distance $l=\gamma
_N\times \tau _N\times c$ from the collision point, where $\tau _N$ is the
lifetime of the excited state in the nucleus rest frame and $c$ is the speed
of light. As an example, for the 4847.2 keV excitation of $^{208}$Pb nucleus
at LHC (which will be considered in section IV) $l$ = 6$\times $10$^{-5}$ m.
Therefore, the detector should be placed close to the collision region. The
5 MeV energy photons emitted in the rest frame of the nucleus will be seen
in the detector as high energy photons with energies up to 30 GeV.

\section{Luminosity Estimations}

In this section we present rough estimations of FEL $\gamma $ - nucleus
luminosities for HERA and LHC based machines. Luminosity of the $\gamma $%
-nucleus collision is given by

\begin{equation}
L=\frac{n_\gamma n_{nuc}}{s_{eff}}f,
\end{equation}
where $n_\gamma $ and $n_{nuc}$ are the number of photons and nuclei in the
FEL and nucleus bunches, respectively, $f=n_b\times f_{rep}$ is the
collision frequency (in which $n_b$ is the number of electron bunches per
pulse and $f_{rep}$ is the repetition rate) and $s_{eff}=4\pi \sigma
_x^{eff}\sigma _y^{eff}$. In the last expression $\sigma $'s denote the size
of the beam on the coordinate axis. $\sigma _x^{eff}$ and $\sigma _y^{eff}$,
on the other hand, are the values which are obtained by choosing the bigger
ones of the corresponding horizontal and vertical sizes of the FEL and the
nucleus beams.

Another parameter which is important in the detector design is the
luminosity per collision which is given by

\begin{equation}
L_0=\frac{n_\gamma n_{nuc}}{s_{eff}}.
\end{equation}
Of course, the bunch structure of linac pulses should be chosen in
accordance with the structure of the nucleus beam.

In the estimations given below we have used HERA $^{12}C$ and LHC $^{208}Pb$
beam parameters \cite{bc...}, which are presented in Table 2. Substituting
the corresponding values from Tables 1 and 2 into Eqs. (2) and (3)
luminosity values given in Table 3 have been obtained. The values $L=5\times
10^{31}cm^{-2}s^{-1}$ for HERA ($^{12}C$) and $L=10^{30}cm^{-2}s^{-1}$ for
LHC ($^{208}Pb$) based FEL $\gamma $ - nucleus colliders seems to be quite
realistic estimations.

\section{Process Examples}

The cross section for the resonant photon scattering is given by the
well-known Breit-Wigner formula:

\begin{equation}
\sigma (\gamma ,\gamma ^{^{\prime }})=\frac \pi {E^2}\times \frac{2J_{exc}+1%
}{2\times (2J_0+1)}\times \frac{B_{in}B_{out}\Gamma ^2}{(E-E_R)^2+\Gamma ^2/4%
}
\end{equation}
where $E$ is the c.m. energy of the incoming photon (in our case it is very
close to that in the rest frame of the nucleus), $J_{exc}$ and $J_0$ are
spins of the excited and ground states of the nucleus, $B_{in}$ and $B_{out}$
are the branching fractions of the excited nucleus into the entrance and
exit channels, respectively, $E_R$ is the energy at the resonance and $%
\Gamma $ is the total width of the excited nucleus. The number of the
scattering events from the nuclei of $^{12}C$ and $^{208}Pb$ have roughly
been calculated below taking their excited states of 4438 keV \cite{C} and
4842.2 keV \cite{swann}\cite{Pb}, respectively. One can easily obtain
corresponding values for other excitations and other nuclei using
appropriate informations from \cite{NDS}.

\subsection{ E = 4438 KeV Excitation of $^{12}C$ at FEL $\gamma $ - HERA}

In this case FEL photons with $\omega _0=8876\ eV$ are needed. According to
Eq. (1) this corresponds to $E_e=6.78\ GeV$ if other parameters of the
undulator are fixed. Using the values of $J_{exc}=2$, $J_0=0$, $\Gamma
\simeq 10.8\times 10^{-3}\ eV$, $B_{in}=B_{out}=1$, the resonant scattering
cross-section has been obtained by Eq. (4) as

\begin{equation}
\sigma ^{res}\simeq 6.2\times 10^{-22}cm^2.
\end{equation}
Taking into account the energy spread of FEL and the nucleus beams ($\Delta
E_\gamma /E_\gamma =\Delta \omega /\omega _0=10^{-3}$ and $\Delta
E_C/E_C=10^{-4}$), the approximate value of averaged cross-section has been
found to be 
\begin{equation}
\sigma ^{av}\simeq \sigma ^{res}\frac \Gamma {\Delta E_\gamma }\simeq
1.5\times 10^{-27}cm^2
\end{equation}
where $\Delta E_\gamma =4438\ eV$. This value corresponds to 6.4$\times $10$%
^9$ events per day for $L=5\times 10^{31}\ cm^{-2}s^{-1}$.

\subsection{ E = 4842.2 KeV Excitation of $^{208}Pb$ at FEL $\gamma $ - LHC}

In this case $\omega _0=1616\ eV$ and $E_e=2.89\ GeV$. Substituting $%
J_{exc}=1$, $J_0=0$, $\Gamma \simeq 5\ eV$ and $B_{in}=B_{out}=1$ into Eq.
(4) the resonant cross-section has been estimated as

\begin{equation}
\sigma ^{res}\simeq 3.1\times 10^{-22}cm^2
\end{equation}
and, consequently, with the same energy spreads given above, the average
cross-section as

\[
\sigma ^{av}\simeq 0.3\times 10^{-24}cm^2 
\]
which corresponds to 2.5$\times $10$^{10}$ events per day for $%
L=10^{30}cm^{-2}s^{-1}$.

\section{Conclusion}

In this study a new method has been presented to investigate the nuclear
excitations. Although the quantitative description was given for only two
nuclei it can easily be extended for others too. The method has the
advantage that the accelerated ionized nuclei see the keV energy FEL photons
as a laser beam in the MeV energy range. In principle, FEL $\gamma $ beam
can be obtained from existing HERA and LEP ring electron beams, too. This
option will be considered elsewhere.

For future investigations the determination of the following nuclear
quantities might be promising: scattering cross-sections, spins and parity
assignments of the excited states.

\bigskip\ 

{\bf ACKNOWLEDGMENTS}

This work is partially supported by Turkish State Planning Organization
under Grant No 97.K.120420.

\bigskip\


\begin{references}
\bibitem{ankara}  Proceedings of the First International Workshop on
Linac-Ring Type ep and $\gamma $p Colliders, 9-11 April 1997, Ankara,
Turkey. To be published in Turkish Journal of Physics.

\bibitem{sultansoy1}  S. Sultansoy ''Four Ways to TeV Scale'' in Ref.[1].

\bibitem{tww}  M. Tigner, B. Wiik and F. Willeke, Proc. 1991 IEEE Particle
Accelerator Conference, Vol.5, p.2910.

\bibitem{acs}  Z. Z. Aydin, A. K. \c {C}ift\c {c}i and S. Sultansoy, Nucl.
Instum. Meth. A351 (1994) 261.

\bibitem{bd}  R. Brinkmann and M. Dohlus, DESY preprint M-95-11, Hamburg
(1995).

\bibitem{bc...}  R. Brinkmann et al., DESY preprint 97-239, Hamburg (1997).

\bibitem{ictp}  S. F. Sultanov, ICTP preprint IC/89/409, Trieste (1989).

\bibitem{csty}  A. K. \c {C}ift\c {c}i, S. Sultansoy, \c {S}. T\"{u}rk\"{o}z
and \"{O}. Yava\c {s}, Nucl. Instrum. Meth. A365 (1995) 317.

\bibitem{ichep}  Z. Z. Aydin et al., Proceedings of the 28'th Int. Conf. on
High Energy Physics, eds Z. Ajduk and A. K. Wroblewski, World Scientific,
Singapore, 1996, p.1752.

\bibitem{FEL}  TESLA-FEL 95-03, June 1995, A VUV Free Electron Laser at The
TESLA Test Facility at DESY.

\bibitem{C}  K. P. Schelhaas et al., Nucl. Phys. A506 (1990) 307.

\bibitem{swann}  C. P. Swann, Phys. Rev. Lett. 32 (1974) 1449.

\bibitem{Pb}  R. M. Laszewski and P. Axel, Phys. Rev. C 19 (1979) 342.

\bibitem{NDS}  Nuclear Data Sheets, Academic Press, Inc.

\newpage\ 

Table 1. TTF FEL parameters

\begin{tabular}{|c|c|}
\hline
Electron beam energy $E_e$, GeV & 1 \\ \hline
Number of electrons per bunch $n_e$, 10$^{10}$ & 0.624 \\ \hline
Pulse length, $\mu $s & 1000 \\ \hline
Number of bunches per pulse $n_b$ & 7200 \\ \hline
Repetition rate, Hz & 10 \\ \hline
Period length of an undulator $l_u$, mm & 27.3 \\ \hline
Peak field in the undulator $B_u$, T & 0.497 \\ \hline
Undulator parameter $k$ & 1.3 \\ \hline
Photon wavelength $\lambda $, nm & 6.4 \\ \hline
Photon energy $\omega _0$, eV & 193 \\ \hline
Number of photons per bunch $n_\gamma $, 10$^{13}$ & 4 \\ \hline
Energy spread of photons $\Delta \omega /\omega _0$, 10$^{-3}$ & 1 \\ \hline
rms beam size $\sigma _{x,y}$, $\mu $m & 50 \\ \hline
\end{tabular}

\medskip\ 

Table 2. Parameters of nucleus beams

\begin{tabular}{|c|c|c|}
\hline
& $^{12}C$ at HERA & $^{208}Pb$ at LHC \\ \hline
Maximum beam energy, TeV & 4.9 & 574 \\ \hline
Particles per bunch $n_{nuc}$, 10$^8$ & 80 & 0.94 \\ \hline
Normalized emittance $\varepsilon ^N$, mm$\times $mrad & 1.25 & 1.4 \\ \hline
Amplitude function at IP $\beta ^{*}$, cm & 20 & 20 \\ \hline
rms beam size at IP $\sigma _{x,y}$, $\mu $m & 22 & 9.7 \\ \hline
Bunch spacing, ns & 192 & 100 \\ \hline
Number of bunches in FEL pulse $n_b$ & 5208 & 10000 \\ \hline
\end{tabular}

\medskip\ 

Table 3. Luminosities of FEL $\gamma $ - nucleus collisions

\begin{tabular}{|c|c|c|}
\hline
& HERA $^{12}C$ & LHC $^{208}Pb$ \\ \hline
Luminosity $L$, 10$^{30}$ cm$^{-2}$s$^{-1}$ & 53 & 1.1 \\ \hline
Luminosity per collision $L_0$, 10$^{26}$ cm$^{-2}$ & 7.2 & 0.17 \\ \hline
\end{tabular}
\end{references}
\end{document}